\documentclass[12pt,aps,onecolumn,showpacs,superscriptaddress]{revtex4}

\usepackage{graphicx}
\usepackage{amsmath}
\usepackage{hyperref}

\newcommand{\widebar}[1]{\ensuremath{\overset{\hspace{0.8mm}\underline{\hspace{2.33mm}}}{#1}}}

\begin{document}

\title{\boldmath Measurement of the $e^+e^- \to \omega\eta$ cross section below
$\sqrt{s}=2$~GeV}

%==============================================
\author{M.~N.~Achasov}
\affiliation{Budker Institute of Nuclear Physics, SB RAS, Novosibirsk 630090, Russia}
\affiliation{Novosibirsk State University, Novosibirsk 630090, Russia}
\author{A.~Yu.~Barnyakov}
\affiliation{Budker Institute of Nuclear Physics, SB RAS, Novosibirsk 630090, Russia}
\affiliation{Novosibirsk State University, Novosibirsk 630090, Russia}
\author{K.~I.~Beloborodov}
\affiliation{Budker Institute of Nuclear Physics, SB RAS, Novosibirsk 630090, Russia}
\affiliation{Novosibirsk State University, Novosibirsk 630090, Russia}
\author{A.~V.~Berdyugin}
\affiliation{Budker Institute of Nuclear Physics, SB RAS, Novosibirsk 630090, Russia}
\affiliation{Novosibirsk State University, Novosibirsk 630090, Russia}
\author{A.~G.~Bogdanchikov}
\affiliation{Budker Institute of Nuclear Physics, SB RAS, Novosibirsk 630090, Russia}
\author{A.~A.~Botov}
\email[e-mail: ]{A.A.Botov@inp.nsk.su}
\affiliation{Budker Institute of Nuclear Physics, SB RAS, Novosibirsk 630090, Russia}
\author{T.~V.~Dimova}
\affiliation{Budker Institute of Nuclear Physics, SB RAS, Novosibirsk 630090, Russia}
\affiliation{Novosibirsk State University, Novosibirsk 630090, Russia}
\author{V.~P.~Druzhinin}
\affiliation{Budker Institute of Nuclear Physics, SB RAS, Novosibirsk 630090, Russia}
\affiliation{Novosibirsk State University, Novosibirsk 630090, Russia}
\author{V.~B.~Golubev}
\affiliation{Budker Institute of Nuclear Physics, SB RAS, Novosibirsk 630090, Russia}
\affiliation{Novosibirsk State University, Novosibirsk 630090, Russia}
\author{L.~V.~Kardapoltsev}
\affiliation{Budker Institute of Nuclear Physics, SB RAS, Novosibirsk 630090, Russia}
\affiliation{Novosibirsk State University, Novosibirsk 630090, Russia}
\author{A.~S.~Kasaev}
\affiliation{Budker Institute of Nuclear Physics, SB RAS, Novosibirsk 630090, Russia}
\author{A.~G.~Kharlamov}
\affiliation{Budker Institute of Nuclear Physics, SB RAS, Novosibirsk 630090, Russia}
\affiliation{Novosibirsk State University, Novosibirsk 630090, Russia}
\author{A.~N.~Kirpotin}
\affiliation{Budker Institute of Nuclear Physics, SB RAS, Novosibirsk 630090, Russia}
\author{D.~P.~Kovrizhin}
\affiliation{Budker Institute of Nuclear Physics, SB RAS, Novosibirsk 630090, Russia}
\affiliation{Novosibirsk State University, Novosibirsk 630090, Russia}
\author{I.~A.~Koop}
\affiliation{Budker Institute of Nuclear Physics, SB RAS, Novosibirsk 630090, Russia}
\affiliation{Novosibirsk State University, Novosibirsk 630090, Russia}
\affiliation{Novosibirsk State Technical University, Novosibirsk 630092, Russia}
\author{A.~A.~Korol}
\affiliation{Budker Institute of Nuclear Physics, SB RAS, Novosibirsk 630090, Russia}
\affiliation{Novosibirsk State University, Novosibirsk 630090, Russia}
\author{S.~V.~Koshuba}
\affiliation{Budker Institute of Nuclear Physics, SB RAS, Novosibirsk 630090, Russia}
\affiliation{Novosibirsk State University, Novosibirsk 630090, Russia}
\author{A.~S.~Kupich}
\affiliation{Budker Institute of Nuclear Physics, SB RAS, Novosibirsk 630090, Russia}
\affiliation{Novosibirsk State University, Novosibirsk 630090, Russia}
\author{K.~A.~Martin}
\affiliation{Budker Institute of Nuclear Physics, SB RAS, Novosibirsk 630090, Russia}
\author{N.~A.~Melnikova}
\affiliation{Budker Institute of Nuclear Physics, SB RAS, Novosibirsk 630090, Russia}
\author{A.~E.~Obrazovsky}
\affiliation{Budker Institute of Nuclear Physics, SB RAS, Novosibirsk 630090, Russia}
\author{E.~V.~Pakhtusova}
\affiliation{Budker Institute of Nuclear Physics, SB RAS, Novosibirsk 630090, Russia}
\author{A.~I.~Senchenko}
\affiliation{Budker Institute of Nuclear Physics, SB RAS, Novosibirsk 630090, Russia}
\author{S.~I.~Serednyakov}
\affiliation{Budker Institute of Nuclear Physics, SB RAS, Novosibirsk 630090, Russia}
\affiliation{Novosibirsk State University, Novosibirsk 630090, Russia}
\author{Z.~K.~Silagadze}
\affiliation{Budker Institute of Nuclear Physics, SB RAS, Novosibirsk 630090, Russia}
\affiliation{Novosibirsk State University, Novosibirsk 630090, Russia}
\author{Yu.~M.~Shatunov}
\affiliation{Budker Institute of Nuclear Physics, SB RAS, Novosibirsk 630090, Russia}
\affiliation{Novosibirsk State University, Novosibirsk 630090, Russia}
\author{D.~A.~Shtol}
\affiliation{Budker Institute of Nuclear Physics, SB RAS, Novosibirsk 630090, Russia}
\affiliation{Novosibirsk State University, Novosibirsk 630090, Russia}
\author{D.~B.~Shwartz}
\affiliation{Budker Institute of Nuclear Physics, SB RAS, Novosibirsk 630090, Russia}
\affiliation{Novosibirsk State University, Novosibirsk 630090, Russia}
\author{A.~N.~Skrinsky}
\affiliation{Budker Institute of Nuclear Physics, SB RAS, Novosibirsk 630090, Russia}
\author{I.~K.~Surin}
\affiliation{Budker Institute of Nuclear Physics, SB RAS, Novosibirsk 630090, Russia}
\affiliation{Novosibirsk State University, Novosibirsk 630090, Russia}
\author{Yu.~A.~Tikhonov}
\affiliation{Budker Institute of Nuclear Physics, SB RAS, Novosibirsk 630090, Russia}
\affiliation{Novosibirsk State University, Novosibirsk 630090, Russia}
\author{A.~V.~Vasiljev}
\affiliation{Budker Institute of Nuclear Physics, SB RAS, Novosibirsk 630090, Russia}
\affiliation{Novosibirsk State University, Novosibirsk 630090, Russia}

\begin{abstract}
The cross section for the process $e^+e^- \to \omega\eta$ is measured in
the center-of-mass energy range 1.34--2.00~GeV. The analysis is based on
data collected with the SND detector at the VEPP-2000 $e^+e^-$ collider. The 
measured $e^+e^- \to \omega\eta$ cross section is the most accurate to date. 
A significant discrepancy is observed between our data and previous BABAR 
measurement.
\end{abstract}

\pacs{13.66.Bc, 14.40.Cs, 14.40.Aq, 13.40.Gp}

\maketitle

%==============================================
\section{Introduction}

The main goal of experiments with the SND detector~\cite{SND} at the $e^+e^-$
collider VEPP-2000~\cite{VEPP} is a precision measurement of the total
cross section of $e^+e^-$ annihilation to the hadrons in the center-of-mass
(c.m.) energy region $E=\sqrt{s}<2$~GeV. The total cross section
is necessary for calculation of the running electromagnetic coupling constant
and the muon anomalous magnetic moment. Below 2~GeV the total 
hadronic cross section is calculated as a sum of exclusive cross sections for
all possible hadronic modes. For some of them, e.g., $\pi^+\pi^-\pi^0\eta$,
$\pi^+\pi^-\pi^0\pi^0\pi^0$, $\pi^+\pi^-\pi^0\pi^0\eta$, which 
may give a sizable contribution to the total hadronic cross section, 
experimental information is scarce or absent. The process 
$e^+e^- \to \pi^+\pi^-\pi^0\eta$ can proceed through the $\omega\eta$, 
$\phi\eta$ intermediate states, the cross sections for which are already 
measured~\cite{OMETA,PHIETA1,PHIETA2}, and also through other states, 
e.g., $\rho a_0(980)$. This work is dedicated to the measurement of the 
$e^+e^- \to \omega\eta$ cross section. We analyze the $\pi^+\pi^-\pi^0\eta$
final state with $\eta$ meson decayed to $\gamma\gamma$. The methods developed
for the selection of $e^+e^- \to \omega\eta \to \pi^+\pi^-\pi^0\eta$ events 
will be used in future detailed study of the process 
$e^+e^- \to \pi^+\pi^-\pi^0\eta$ and measurement its total cross section.

Previously, the process $e^+e^- \to \omega\eta$ was measured using the ISR 
technique in the BABAR experiment~\cite{OMETA} in the six pion final state.

%==============================================
\section{\label{DATA}Detector and experiment}

SND~\cite{SND} is a general purpose nonmagnetic detector. Its main part is a
spherical three-layer electromagnetic calorimeter with 560 individual NaI(Tl)
crystals per layer. The calorimeter covers a solid angle of 95\% of $4\pi$ and
has a width of $13.4X_0$, where $X_0$ is a radiative length. The calorimeter 
energy resolution for photons is 
$\sigma_E/E_\gamma = 4.2\%/\sqrt[4]{E_\gamma(\text{GeV})}$. The angular
resolution is about $1.5^\circ$. 
Inside the calorimeter, around the collider   beam pipe, a tracking 
system is located, which consists of a nine-layer drift chamber and 
a proportional chamber with cathode-strip readout in a common gas volume.
The tracking system covers a solid angle of 94\% of $4\pi$. 
Its angular resolution is $0.45^\circ$ in the azimuthal angle and 
$0.8^\circ$ in the polar angle. A system of aerogel Cherenkov counters 
located between the tracking system and the calorimeter is used for
charged kaon identification. Outside the calorimeter, a muon detector 
consisting of proportional tubes and scintillation counters is placed. 

The analysis is based on data with an integrated luminosity of 27~pb$^{-1}$
recorded with the SND detector in 2011--2012 in 36 energy points 
above the threshold of the process under study.

During the experiment, the beam energy was determined using measurements 
of the magnetic filed in the collider bending magnets. To fix the absolute 
energy scale, a scan of the $\phi(1020)$ resonance was performed and 
its mass was measured. In 2012 the beam energy was measured in
several energy points near 2~GeV by
the back-scattering-laser-light system~\cite{COMPTON1,COMPTON2}.
The absolute energy measurements were used for calibration of the
momentum measurement in the CMD-3 detector, which collected data at
VEPP-2000 simultaneously with SND. 
The absolute c.m.\ energies for all scan points were then 
determined using average momentum in Bhabha and $e^+e^- \to p\bar{p}$ events
with accuracy of 2--6~MeV~\cite{BEAM1}.
Because of the absence of any narrow structures in  
the $e^+e^- \to \omega\eta$ cross section, the 36 energy points are merged 
into 13 energy intervals listed in Table~\ref{TAB:CRS}. For each interval
the weighted average value of the c.m.\ energy ($\widebar{E}_i$) is also
listed, which is calculated as 
$\sum E_j L_j \sigma_{vis}(E_j)/\sum L_j \sigma_{vis}(E_j)$, where the sum
is over the scan energy points included into the $i$-th interval,
$L_j$ is the integrated luminosity for the $j$-th scan point, and
$\sigma_{vis}$ is the visible cross section for $e^+e^- \to \omega\eta$
defined in Sec.~\ref{CRS}.

Simulation of the signal and background processes is done with Monte Carlo (MC)
event generators. The generators take into account 
radiative corrections to the initial particles calculated according 
to Ref.~\cite{KURAEV}. The angular distribution of additional photons
radiated by the initial electron and positron is simulated
according to Ref.~\cite{BONNEAU}. The cross-section energy dependences
needed for radiative-correction calculation are taken from existing data,
e.g., from Ref.~\cite{OMETA} for the process $e^+e^- \to \omega\eta$.

Interactions of the generated particles with the detector materials are
simulated using GEANT4 software~\cite{GEANT4}. The simulation takes into 
account variation of experimental conditions during data taking, 
in particular dead detector channels and beam-induced background. The beam 
background leads to appearance of spurious photons and charged particles in
detected events. To take this effect into account, simulation
uses special background events recorded during data taking with
a random trigger, which are superimposed on simulated events.

%==============================================
\section{\label{LUMINOSITY}Luminosity measurement}

The process of Bhabha scattering $e^+e^- \to e^+e^-$ is used for
luminosity measurement. Bhabha events are selected with the following 
conditions. They must contain at least two charged particles originated 
from the beam interaction region. The particle energies are determined 
on their energy depositions in the calorimeter. 
Further conditions are applied to the two most energetic charged
particles. Their energies must be greater than $0.6E_{beam}$, where $E_{beam}$
is the beam energy, and their polar $(\theta_{1,2})$ and azimuthal $(\phi_{1,2})$
angles must satisfy the conditions 
$(180^\circ-|\theta_1-\theta_2|)/2>50^\circ$,
$|\theta_1+\theta_2-180^\circ|<15^\circ$, 
$||\phi_1-\phi_2|-180^\circ|<10^\circ$.

The detection efficiency and cross section for Bhabha events 
are determined using the event generator BHWIDE~\cite{BHWIDE}.
The integrated luminosity measured for each
energy interval is listed in Table~\ref{TAB:CRS}. The theoretical
uncertainty of the luminosity measurement is less than 0.5\%. 
The systematic uncertainty on the detection efficiency is estimated
by variation of the selection criteria used and does not exceed 2\%.

%==============================================
\section{\label{SEL}Event selection}

At the first stage of analysis, events with two or three charged particles
originated from the interaction region, and at least four photons with
energy greater than 20~MeV are selected. The total energy deposition
in the calorimeter for these events is required to be greater than 300~MeV.

For selected events the vertex fit is performed using parameters
of two charged tracks. The quality of the vertex fit is characterized by the 
parameter $\chi^2_r$. If there are three charged tracks in
an event, two of them with the lowest $\chi^2_r$ value are selected. The found
vertex is used to refine the parameters of charged particles and photons.
Then the kinematic fit to the $e^+e^- \to \pi^+\pi^-\pi^0\gamma\gamma$ 
hypothesis is performed with the requirement of energy and momentum 
balance and the $\pi^0$ mass constraint. The $\pi^0$ candidate is a two 
photon pair with invariant mass in the range $70 < m_{12} < 200$~MeV. 
The invariant mass of the second photon pair ($\eta$-meson candidate) must 
be in the range $400 < m_{34} < 700$~MeV. The quality of the kinematic fit 
is characterized by the parameter $\chi^2_{3\pi2\gamma}$. All possible
combinations of photons are tested, and the combination with the smallest
$\chi^2_{3\pi2\gamma}$ value is chosen. The photon parameters after
the kinematic fit are used to recalculate the $\eta$-candidate invariant mass 
($M_\eta$). The event is then refitted with $\eta$-mass constraint.
The refined energy of the $\eta$-meson candidate is used to calculate the 
invariant mass of the system recoiling against $\eta$ meson ($M_\eta^{rec}$).

Events of the process $e^+e^- \to \omega\eta$ are selected by the
conditions $\chi^2_{3\pi2\gamma} < 30$ and $0.65 < M_\eta^{rec} < 0.9$~GeV. 
The main background source is the process $e^+e^- \to \pi^+\pi^-\pi^0\pi^0$. 
For its suppression, a kinematic fit to the hypothesis $e^+e^- \to
\pi^+\pi^-\pi^0\pi^0(\gamma)$ is performed. In this fit, radiation
of an additional photon along the beam axis is allowed.  Events with
$\chi^2_{4\pi(\gamma)} < 200$ are rejected.

\begin{figure}
\includegraphics[width=0.45\textwidth]{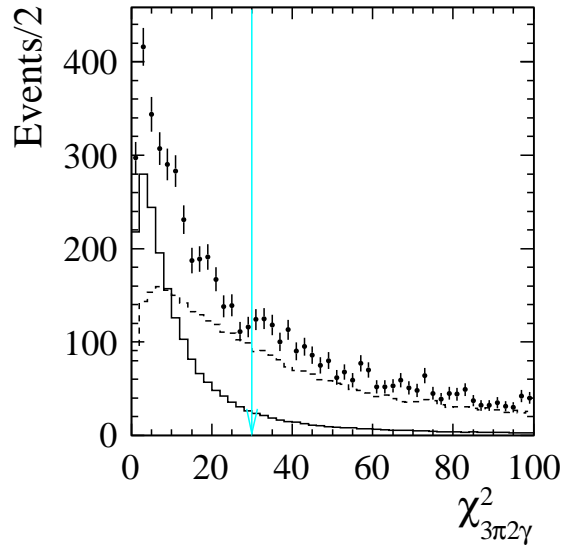}
\caption{\label{FIG:CHI2IET}The $\chi^2_{3\pi2\gamma}$ distribution for selected data events  
(points with error bars). The solid and dashed histograms represent the
shapes of signal $e^+e^- \to \omega\eta$ and background 
$e^+e^- \to \pi^+\pi^-\pi^0\pi^0$ distributions obtained using MC simulation,
respectively. The arrow indicate the boundary of the condition
$\chi^2_{3\pi2\gamma} < 30$.}
\end{figure}

The $\chi^2_{3\pi2\gamma}$ distribution for selected data events is shown in
Fig.~\ref{FIG:CHI2IET} in comparison with the simulated distributions 
for signal $e^+e^- \to \omega\eta$ and background 
$e^+e^- \to \pi^+\pi^-\pi^0\pi^0$ events. The narrow signal peak near zero
is clearly seen in the distribution. However, the region 
$\chi^2_{3\pi2\gamma}<30$ contains a significant amount of background events. 

%==============================================
\section{\label{FIT}Determination of the number of signal events}

\begin{figure}
\includegraphics[width=0.47\textwidth]{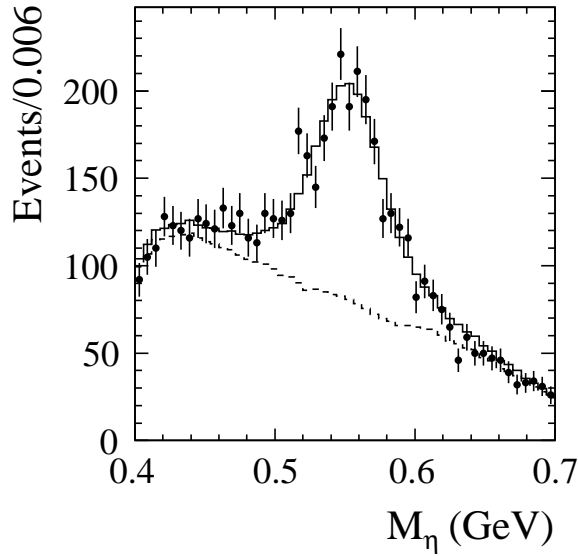}
\caption{\label{FIG:META}The spectrum of the two-photon invariant mass of
the $\eta$-meson candidate for selected data events (points with error bars).
The solid histogram is the result of the fit to the data spectrum
with a sum of signal and background distributions.
The background contribution is shown by the dashed histogram.}
\end{figure}

The $M_\eta$ spectrum for selected data events is shown in 
Fig.~\ref{FIG:META}. It is seen that only about 25\% of events 
contain an $\eta$ meson.
The spectrum is fitted with a sum of signal and background distributions.
The background distribution is obtained using simulation of  
the process $e^+e^- \to \pi^+\pi^-\pi^0\pi^0$.
A possible background simulation inaccuracy is taken into account by 
introducing a scale factor $\alpha_{4\pi}$. For energies below 1.594~GeV,
the value of $\alpha_{4\pi}$ found in the fit is consistent with unity.
At higher energies, there is significant background contribution
from other processes, e.g., $e^+e^- \to \pi^+\pi^-\pi^0\pi^0\pi^0$.
In this region $\alpha_{4\pi}$ is fixed to unity, and a linear function 
is added to describe contribution of other background processes.
It is worth to note that in the energy region above 1.594~GeV 
the shape of the $M_\eta$ distribution for $e^+e^- \to \pi^+\pi^-\pi^0\pi^0$ 
events is close to linear.

The signal distribution is described by a sum of three Gaussian distributions
with parameters determined from the fit to the $M_\eta$ distributions for 
$e^+e^- \to \omega\eta \to \pi^+\pi^-\pi^0\eta$ simulated events.
To account for a possible inaccuracy of the signal simulation, two parameters
are introduced: mass shift $\Delta M_\eta$ and smearing parameter 
$\Delta \sigma_{M_\eta}$. The latter is quadratically added to all Gaussian 
sigmas. The parameters $\Delta M_\eta$ and $\Delta \sigma_{M_\eta}$ are
determined from the fit to the $M_\eta$ spectrum for events from
the energy range $1.544 \leq E < 1.794$~GeV, where the signal-to-background 
ratio is maximal. The obtained values $\Delta M_\eta =
-0.2\pm1.3$~MeV and $\Delta\sigma_{M_\eta} = 8.0\pm7.6$~MeV are consistent
with zero. Therefore, these parameters are set to zero in the fit.
Their errors are used to estimate systematic uncertainty 
in the fitted number of signal events due to a possible difference 
between data and simulation in the $\eta$-meson line shape. This 
uncertainty is found to be 1.6\%.

To estimate the systematic uncertainty due to imperfect description of
the shape of the background distribution, we perform the fit with
an additional linear background as described above below 1.594~GeV, and
without the linear background but with free $\alpha_{4\pi}$ above.
The obtained systematic uncertainty is 5\% below 1.594~GeV and 1.2\% above.

\begin{figure}
\includegraphics[width=0.47\textwidth]{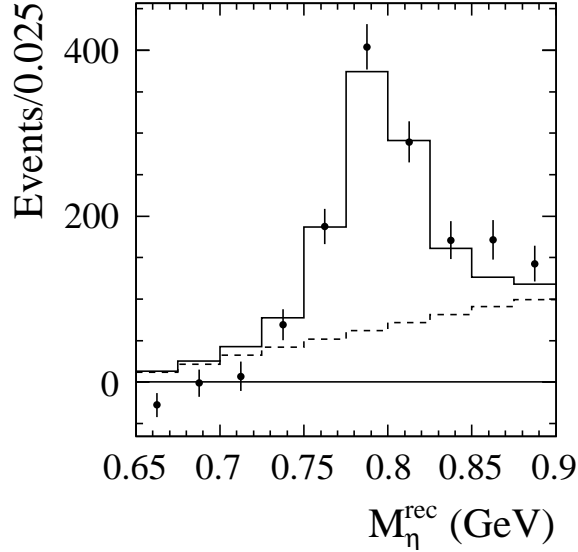}
\caption{\label{FIG:OMEGA}The $M_\eta^{rec}$ distribution for data 
$e^+e^- \to \pi^+\pi^-\pi^0\eta$ events (points with error bars). The solid 
histogram represents the result of the fit described in the text. The dashed 
histogram shows the fitted distribution for non-$\omega\eta$ events.}
\end{figure}

The number of signal and background events obtained from the fit to the 
$M_\eta$ spectrum in Fig.~\ref{FIG:META} are $1413 \pm 67$ and
$4123 \pm 88$, respectively. The events with $\eta$ meson
belong to the process $e^+e^- \to \pi^+\pi^-\pi^0\eta$. To
obtain number of $e^+e^- \to \omega\eta$ events, the $M_{\eta}^{rec}$ spectrum
is analyzed. We divide the interval $0.65<M_{\eta}^{rec}<0.9$~GeV 
into 10 subintervals. In each subinterval the number of events with $\eta$ 
meson is determined from the fit to the $M_\eta$ distribution. The results
of these fits are shown in Fig.~\ref{FIG:OMEGA} as a $M_\eta^{rec}$ histogram.

Such $M_\eta^{rec}$ distributions are obtained for each c.m.\ energy interval
listed in Table~\ref{TAB:CRS}. The $M_\eta^{rec}$ distributions are fitted 
with a sum of the distribution for the process $e^+e^- \to \omega\eta$, 
which has a peak near the $\omega$-meson mass, and a flat distribution
for non-$\omega\eta$ events. To obtain the shape of the latter distribution,
the $e^+e^- \to \pi^+\pi^-\pi^0\eta$ simulation with the uniform 
distribution of the final particle momenta over the phase space is used.
Above 1.594~GeV, where the simulated distribution is close to linear,
a linear function is used in the fit to describe the non-$\omega\eta$ 
background. It should be noted that the fitted numbers of background events 
are consisted with zero below 1.594~GeV.

Our preliminary analysis of intermediate states in the process 
$e^+e^- \to \pi^+\pi^-\pi^0\eta$~\cite{MESON2016} shows that a significant 
contribution to the cross section comes from the $\rho a_0(980)$ 
intermediate state. Therefore, at energies above 1.594~GeV, the alternative 
background model $e^+e^- \to \rho a_0(980)$ is also tested. The difference 
between fit results for the two background models is used as
an estimate of the systematic uncertainty. Below 1.594~GeV,
where the non-$\omega\eta$ background is small,
the difference between fit results with nonzero and zero background 
is taken as an estimate of the systematic uncertainty associated
with the shape of the non-$\omega\eta$ distribution. 

The $\omega$-meson line shape is described by a sum of three Gaussian 
distributions with parameters obtained from the fit to the 
the simulated signal $M_\eta^{rec}$ distribution.
To obtain corrections for the data-simulation difference in the mass
scale and mass resolution, the parameters $\Delta M_\omega$ and
$\Delta \sigma_\omega$ are introduced.
These parameters are determined from the fit to the total mass spectrum
shown in Fig.~\ref{FIG:OMEGA}. The $\Delta \sigma_\omega$ is found to
be consistent with zero, while $\Delta M_\omega = 7.5 \pm 1.9$~MeV.
The systematic uncertainty due to imperfect simulation
the $\omega$-meson line shape is estimated to be 1.4\%.
The total number of $e^+e^- \to \omega\eta$ events obtained from the fit
to the $M_\eta^{rec}$ distribution shown in Fig.~\ref{FIG:OMEGA} is 
$852 \pm 69$. The number of background events is $564 \pm 80$.

The obtained numbers of $e^+e^- \to \omega\eta$ events 
for different energy intervals are listed in Table~\ref{TAB:CRS}. 
The first error is statistical and the second is
systematic, due to the background description in the fit to the
$M_\eta^{rec}$ distribution. The energy independent systematic uncertainty 
on the number of signal events is 5.4\% below 1.594~GeV and 2.4\% above.

%==============================================
\section{Detection efficiency}

\begin{figure}
\includegraphics[width=0.47\textwidth]{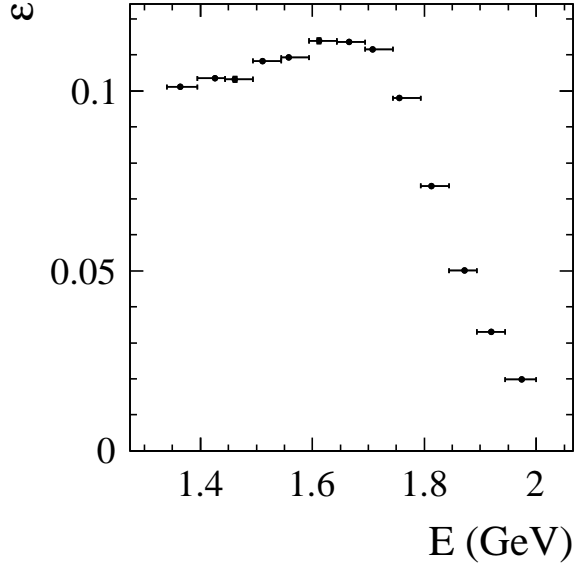}
\caption{\label{FIG:EFF}The energy dependence of the detection efficiency for 
the process $e^+e^- \to \omega\eta \to \pi^+\pi^-\pi^0\eta$.}
\end{figure}

The detection efficiency for events of the process 
$e^+e^- \to \omega\eta \to \pi^+\pi^-\pi^0\eta$ is 
determined using MC simulation. As was discussed in Sec.~\ref{DATA}, the
simulation takes into account radiative corrections.
Therefore, the detection efficiency depends on the Born cross section 
used in simulation. For the process under study, the Born cross section 
measured in the work~\cite{OMETA} is taken as a first approximation. 
Then the cross section is corrected taking into account our measurement,
and the detection efficiency is recalculated. Then the third iteration 
is performed. The energy dependence of the
detection efficiency obtained is shown in Fig.~\ref{FIG:EFF}. The
efficiency decrease above 1.7~GeV is explained by the steep falloff of
the $e^+e^- \to \omega\eta$ cross section in this energy region and increase 
of the fraction of events with a hard photon radiated from the initial state.
The difference between the detection efficiencies found after the second and 
third iterations is taken as an estimate of the model uncertainty.
It is 1\% below 1.694~GeV and 10\% above.

Imperfect simulation of detector response leads to a difference between the 
actual detection efficiency $\varepsilon$ and the efficiency determined
using MC simulation $\varepsilon_{MC}$:
\begin{equation}
\varepsilon = \varepsilon_{MC}\prod_{i=1}^{n}(1+\delta_i),
\end{equation}
where $\delta_i$ are the efficiency corrections for different effects.
The main selection criterion for signal events 
is $\chi^2_{3\pi2\gamma} < 30$. The quality of the simulation of the
kinematic-fit $\chi^2$ distribution is studied using events of the 
process $e^+e^- \to \omega\pi^0  \to \pi^+\pi^-\pi^0\pi^0$, 
which has a large cross section and the same number of the final particles
as the process under study. Events 
from the energy range $1.394 \leq E < 1.594$~GeV, where the $e^+e^- \to \omega\pi^0$ 
cross section is maximal, are selected using the preliminary conditions
described in Sec.~\ref{SEL}, and kinematically fitted to the 
$e^+e^- \to \pi^+\pi^-\pi^0\pi^0$ hypothesis.
The efficiency correction is calculated as 
\begin{equation}
\delta_1 = \frac{N'_{MC}/N_{MC}}{N'/N}-1,
\label{EQ:EFFCORR}
\end{equation}
where $N$ and $N_{MC}$ are the numbers of signal events selected with 
the standard criteria in data and simulation, while $N'$ and 
$N'_{MC}$ are the numbers of events selected with a looser condition 
on the parameter under study. To determine the number of 
$e^+e^- \to \omega\pi^0$ events, the spectrum of the invariant mass recoiling
against most energetic $\pi^0$ meson in an event, which has a peak
at the $\omega$-meson mass, is fitted. From the numbers of events in the
$\omega$ peak obtained with the conditions $\chi^2_{4\pi} < 30$ and 
$\chi^2_{4\pi} < 200$, the correction value is found to be
$\delta_1 = (2.5 \pm 1.1)$\%.

To determine the correction for the condition $\chi^2_{4\pi(\gamma)} > 200$,
events from the energy region $1.594 \leq E < 1.744$~GeV selected with the tighter 
cuts $\chi^2_{3\pi2\gamma} < 20$ and $0.76 < M_\eta^{rec} <0.83$~GeV are used.
The numbers of $\eta$-meson events with and without  the condition on
$\chi^2_{4\pi(\gamma)}$ are obtained from the fit to the  $M_\eta$
distribution. The correction is calculated to be $\delta_2 = (3.8 \pm 4.6)$\%.

The difference between data and simulation in photon conversion in 
detector material before the tracking system is studied
using events of the process $e^+e^- \to \gamma\gamma$. The corresponding 
efficiency correction is $\delta_3 = (-1.35 \pm 0.05)$\%.

The largest part of the systematic uncertainties associated with 
data-MC simulation difference in track reconstruction cancels as a result 
of luminosity normalization. The difference in the track reconstruction for
electrons and pions was studied in Ref.~\cite{TRACKS}. The corresponding
correction $\delta_4 = (-0.3 \pm 0.2)$\%.

The total correction is $(4.7 \pm 4.7)$\%. The corrected values of 
the detection efficiency are listed in Table~\ref{TAB:CRS}.
The statistical error of the efficiency is less than 1\% and included to
the statistical error of the measured cross section.

%==============================================
\section{\label{CRS}The Born cross section for
\texorpdfstring{\boldmath $e^+e^- \to \omega\eta$}{e+e- -> omega eta}}
\begin{table*}
\centering
\caption{\label{TAB:CRS}The c.m.\ energy interval, weighted average energy for the interval 
($\protect\widebar{E}$), integrated luminosity ($L$), number of selected data events ($N$),
detection efficiency ($\varepsilon$), radiative correction factor ($1+\delta$),
measured Born cross section ($\sigma$). For $N$ and $\sigma$, the statistical and 
energy dependent systematic errors are quoted. The energy independent 
uncertainty on the cross section is 7.5\%, 5.8\%, and 11.5\% in the energy 
ranges $E<1.594$~GeV, $1.594 \leq E < 1.694$~GeV, $E \geq 1.694$~GeV, respectively.}
\begin{tabular*}{\textwidth}{c@{\extracolsep{\fill}}cccccc}
\hline
\hline
Energy interval (GeV) & $\widebar{E}$ (GeV)& $L$ (nb$^{-1}$)& $N$ & $\varepsilon$ (\%)& $1+\delta$ & $\sigma$ (nb)\\
\hline
1.340--1.394 & 1.36 & 2082 & $-10\pm7\pm0$	  & 10.1 & 0.78	& $-0.07\pm0.05\pm0$\\
1.394--1.444 & 1.43 & 2256 & $32\pm11\pm0$	  & 10.4 & 0.83	& $0.19\pm0.06\pm0$\\
1.444--1.494 & 1.46 & 1095 & $30\pm8\pm0$	  & 10.3 & 0.85	& $0.35\pm0.10\pm0$\\
1.494--1.544 & 1.51 & 2193 & $28\pm18^{+36}_{-0}$ & 10.8 & 0.87	& $0.15\pm0.10^{+0.19}_{-0}$\\
1.544--1.594 & 1.56 & 1024 & $76\pm10\pm0$	  & 10.9 & 0.87	& $0.87\pm0.11\pm0.01$\\
1.594--1.644 & 1.61 & 1008 & $111\pm19^{+4}_{-0}$ & 11.4 & 0.86	& $1.26\pm0.21^{+0.05}_{-0.01}$\\
1.644--1.694 & 1.67 & 1854 & $338\pm33^{+16}_{-0}$& 11.4 & 0.89	& $2.01\pm0.20^{+0.10}_{-0.03}$\\
1.694--1.744 & 1.71 & 1540 & $140\pm27^{+23}_{-0}$& 11.2 & 1.05	& $0.87\pm0.18^{+0.15}_{-0.02}$\\
1.744--1.794 & 1.76 & 1722 & $88\pm25^{+3}_{-0}$  & 9.8  & 1.31	& $0.44\pm0.17^{+0.04}_{-0.03}$\\
1.794--1.844 & 1.81 & 2927 & $55\pm25^{+2}_{-0}$  & 7.4	 & 1.71 & $0.17\pm0.14^{+0.03}_{-0.02}$\\
1.844--1.894 & 1.87 & 2678 & $-7\pm19^{+4}_{-0}$  & 5.0	 & 2.21 & $-0.03\pm0.15^{+0.02}_{-0}$\\
1.894--1.944 & 1.92 & 3702 & $-17\pm17^{+5}_{-0}$ & 3.3	 & 2.29 & $-0.06\pm0.14^{+0.03}_{-0.01}$\\
1.944--2.000 & 1.97 & 2930 & $-11\pm8^{+2}_{-0}$  & 2.0	 & 3.30 & $-0.06\pm0.17^{+0.03}_{-0.01}$\\
\hline		  
\hline
\end{tabular*}
\end{table*}
The experimental values of the $e^+e^- \to \omega\eta$ visible cross section 
are calculated as follows
\begin{equation}
  \sigma_{vis,i} = \frac{N_i}{L_i \varepsilon_i B(\omega\to\pi^+\pi^-\pi^0)},
\end{equation}
where $N_i$, $L_i$, and $\varepsilon_i$ are the number of selected data 
events, integrated luminosity, and detection efficiency for the $i$-th energy
interval, and $B(\omega \to \pi^+\pi^-\pi^0)$ is the branching fraction for the
$\omega \to \pi^+\pi^-\pi^0$ decay.

The visible cross section is related to the Born cross section ($\sigma$) by
the following expression~\cite{KURAEV}:
\begin{equation}
  \sigma_{vis}(E) = \int_0^{x_{max}} F(x,E)\sigma(E\sqrt{1-x})dx,
  \label{EQ:CRS_VIS}
\end{equation}
where the function $F(x,E)$ describes the probability of radiation of
photons with total energy $xE/2$ by the initial electron and positron. 
The right side of Eq.~(\ref{EQ:CRS_VIS}) can be rewritten in 
the more conventional form:
\begin{equation}
  \int_0^{x_{max}} F(x,E)\sigma(E\sqrt{1-x})dx = \sigma(E)(1+\delta(E)),
  \label{EQ:CRS_VIS2}
\end{equation}
where $\delta(E)$ is the radiative correction.

Experimental values of the Born cross section are determined
as follows. The energy dependence of the measured visible cross section 
is fitted with Eq.~(\ref{EQ:CRS_VIS}), in which the Born cross section 
is given by a theoretical model describing data well. The model parameters
obtained in the fit are used to calculate $\delta(\widebar{E}_i)$, where 
$\widebar{E}_i$ is the weighted average c.m.\ energy for $i$-th energy interval, 
defined in Sec.~\ref{DATA}. The values of the Born cross 
section are then obtained as 
$\sigma_i = {\sigma_{vis,i}}/(1+\delta(\widebar{E}_i))$.

The Born cross section is described by a sum of two resonance contributions:
\begin{widetext}
\begin{equation}
   \sigma(E) = \frac{12\pi}{E^3}\left|
   \sqrt{\frac{B_{\omega'}}
   {P_f(m_{\omega'})}}\frac{m_{\omega'}^{3/2}\Gamma_{\omega'}}{D_{\omega'}}-
   \sqrt{\frac{B_{\omega''}}
   {P_f(m_{\omega''})}}\frac{m_{\omega''}^{3/2}\Gamma_{\omega''}}{D_{\omega''}}
   \right|^2P_f(E),
   \label{EQ:VMD}
\end{equation}
\end{widetext}
where $B_V = B(V \to e^+e^-)B(V \to \omega\eta)$ is
the product of the branching fractions for the $V$ decay to 
$e^+e^-$ and $\omega\eta$, $D_V = E^2-m_V^2+iE\Gamma_V$, $m_V$ and $\Gamma_V$
are the mass and width of the resonance $V$ ($V=\omega'$ or $\omega''$).
The phase space factor $P_f(E)$ is given by
\begin{widetext}
\begin{equation}
  P_f(E) = q(E)^3,\, q(E) = \frac{1}{2E}\sqrt{(E^2-(m_\omega-m_\eta)^2)(E^2-(m_\omega+m_\eta)^2)}.
\end{equation}
\end{widetext}
The first term in Eq.~(\ref{EQ:VMD}) describes the $\omega(1420)$ contribution. 
The second term is a sum of contributions of the $\omega(1650)$ and
$\phi(1680)$ resonances. The phase between the first and second terms is
chosen to be equal to $\pi$ (see discussion below).

\begin{figure}
\includegraphics[width=0.47\textwidth]{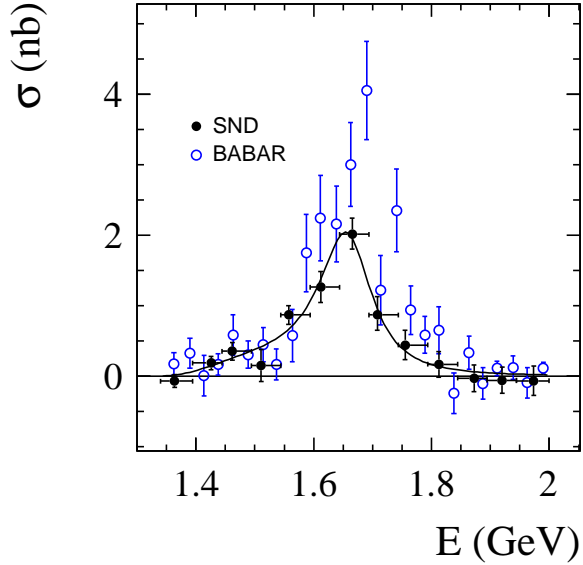}
\caption{\label{FIG:CRS}The $e^+e^- \to \omega \eta$ cross section measured in this work 
(filled circles) and in the BABAR experiment~\cite{OMETA} (open circles).
The curve is the result of the fit described in the text. The errors of
the SND data are statistical and systematic, combined in quadrature.}
\end{figure}

The free fit parameters are $B_{\omega'}, B_{\omega''},
m_{\omega''}, \Gamma_{\omega''}$. The mass and width
of the $\omega(1420)$ resonance are fixed at their Particle Data Group
(PDG) values~\cite{PDG}.
In the fit negative numbers of events (see Table~\ref{TAB:CRS}) are
substituted by zero values. The fitted parameters are listed in 
Table~\ref{TAB:APPROX}. The fit yields $\chi^2=14.5$ for 9 degrees of
freedom. The fitted curve together with obtained values of the Born cross
section is shown in Fig.~\ref{FIG:CRS}. The fit performed with zero
phase between the $\omega'$ and $\omega''$ 
amplitudes provides a significantly worse ($\chi^2=41.6$) description 
of the cross-section data.
\begin{table}
\centering
\caption{\label{TAB:APPROX}The obtained fit parameters.}
\begin{tabular*}{0.35\textwidth}{l@{\extracolsep{\fill}}c}
\hline
\hline
$B_{\omega'}\times10^{7}$ & $0.16^{+0.09}_{-0.07}$\\
$B_{\omega''}\times10^{7}$ & $4.4\pm 0.5$\\
$M_{\omega''}$ (MeV) &  $1660\pm10$\\
$\Gamma_{\omega''}$ (MeV) & $110\pm20$\\
\hline
\hline
\end{tabular*}
\end{table}

The fitted $\omega''$ mass is in agreement with the PDG mass
of both $\omega(1650)$ and $\phi(1680)$ resonances~\cite{PDG},
while the fitted width is smaller than the PDG
estimate for the $\omega(1650)$ width, $315\pm 35$~MeV~\cite{PDG}, but 
agrees with the PDG value, $150\pm 50$~MeV~\cite{PDG}, for the
$\phi(1680)$ resonance. The contribution of the $\omega(1420)$ is
small compared with that of the $\omega''$. However, 
this contribution is
necessary to describe the asymmetry of the peak in the measured cross 
section. The asymmetry is explained by constructive interference
of the $\omega'$ and $\omega''$ amplitudes on 
the left side of the peak and destructive interference on the right
side.

The numerical values of the measured Born cross section and 
radiative correction are listed in Table~\ref{TAB:CRS}. 
The quoted errors of the cross section are statistical and 
energy dependent systematic. The latter includes the energy dependent 
systematic uncertainty on the number of events and the model 
uncertainty of the radiative correction, which is estimated by
varying the fitted parameters within their errors. The sources
of the energy independent systematic uncertainty are listed
in Table~\ref{TAB:SYS}. This uncertainty is equal to
7.5\%, 5.8\%, and 11.5\% in the energy ranges 
$E<1.594$~GeV, $1.594\leq E<1.694$~GeV, $E\geq 1.694$~GeV, respectively.
\begin{table*}
\centering
\caption{\label{TAB:SYS}The systematic uncertainties on the measured cross 
section.}
\begin{tabular*}{0.8\textwidth}{l@{\extracolsep{\fill}}l}
\hline
\hline
Source & Value (\%)\\
\hline
Luminosity & 2\\
Uncertainties on $\Delta M_\eta$, $\Delta \sigma_{M_\eta}$ & 1.6\\

Background shape in the $M_\eta$ distribution & 5.0 at $E < 1.594$~GeV\\
                                              & 1.2 at $E \geq 1.594$~GeV\\
Uncertainty on $\Delta M_\omega$ & 1.4\\

Model uncertainty of the detection efficiency & 1 at $E < 1.694$~GeV\\
                                              & 10 at $E \geq 1.694$~GeV\\     
Condition on $\chi^2_{3\pi2\gamma}$ & 1.1\\
Condition on $\chi^2_{4\pi}$ & 4.6\\
Photon conversion & 0.05\\
Charged track reconstruction & 0.2\\
\hline
Total  & 7.5 at $E < 1.594$~GeV\\
       & 5.8 at $1.594 \leq E < 1.694$~GeV\\
       & 11.5 at $E \geq 1.694$~GeV\\
\hline
\hline
\end{tabular*}
\end{table*}

The comparison of our cross section data with the previous BABAR 
measurement~\cite{OMETA}
is shown in Fig.~\ref{FIG:CRS}. Our results have better accuracy and 
disagree with the BABAR data at $E >1.6$~GeV.

\section{Summary}
In this paper we have analyzed data collected with the SND detector at
the VEPP-2000 $e^+e^-$ collider. In the $\pi^+\pi^-\pi^0\eta$ final state 
we have selected about 850 $\omega\eta$ events and have measured the
$e^+e^- \to \omega\eta$ cross section in the c.m.\ energy range 1.34--2.00~GeV.
The obtained cross section data are the most accurate to date. Above
1.6~GeV they disagree with previous BABAR measurements~\cite{OMETA}.
The measured cross section is well fitted by a sum of two resonance
contributions, from the $\omega(1420)$ and from an effective resonance 
describing the $\omega(1650)$ and $\phi(1680)$ contributions. The fitted 
$\omega(1420)$ amplitude is small, but necessary to describe the asymmetry 
of the peak in the measured cross section. 

\section{ACKNOWLEDGMENTS}
Part of this work related to the photon reconstruction algorithm in the
electromagnetic calorimeter is supported
by the Russian Science Foundation (project No.~14-50-00080).

%==============================================

\end{document}